\documentclass[a4paper,12pt,oneside]{article}
\usepackage[english]{babel}
\usepackage{graphicx}

\begin{document}

\title{ Magnetic field of young star RW Aur}

\author{A.\,V. Dodin$^1$, S.\,A. Lamzin$^1$, G.\,A. Chountonov$^2$}
\date{ \it \small
1) Sternberg Astronomical Institute, Moscow, 119992, Russia\footnote{
{\it Send offprint requests to}: A. Dodin e-mail: samsebedodin@gmail.com}\\
2) Special Astrophysical Observatory of the Russian AS, Nizhnij Arkhyz
369167, Russia\\
}

\maketitle

\bigskip



\bigskip

%%%%%%%%%%%%%%%%%%%%%%%%%%%%%%%%%%%%%%%%%%%%%%%%%%%%%%

\section*{Abstract}

  Results of longitudinal magnetic field $B_z$ measurements for young star
RW Aur A are presented. We found that $B_z$ in the formation region of
He\,I 5876 line's narrow component varies from $-1.47 \pm 0.15$ kG to $+1.10
\pm 0.15$ kG. Our data are consistent with a stellar rotational period of
$\simeq 5.6^d$ and with a model of two hotspots with an opposite polarity of
magnetic field and with a difference in a longitude about $180^o.$
The spot with $B_z<0$ is located at the hemisphere above the midplane of
RW Aur's accretion disc and the spot with $B_z>0$ is below the midplane.

  The following upper limits for $B_z$ (at $3\sigma$ level) were found
after averaging of all our observations: 180 G for photosperic lines,
220 G and 230 G for formation regions of H$_\alpha$ and [OI] 6300 lines
respectively. Upper limit 600 G were found in the region where
broad components of emisson lines form.

 For two cases out of 11 we observed the field in a formation region of a
blue absorption wing of Na\,I D lines i.e. in an outflow: $B_z= -180 \pm
50$ G and $-810 \pm 80$ G.

   Radial velocity of RW Aur's photospheric lines averaged over all our
observations is $\simeq + 10.5$ km\,s$^{-1}$ what is 5.5 km\,s$^{-1}$ less
than value derived ten years earlier by Petrov et al. (2001). In this
connection, we discuss a possibility that RW Aur is not a binary but a
triple system.

			%%%%%%%%%%%%%%%%%%%%%%%%%%%%%%%%%

\section*{Introduction}

  T Tauri Stars are young $(t<10^{7}$ yr) low mass $(M\le 2\,M_\odot)$ stars
at a stage of a gravitational contraction towards the main sequence.
Magnetic fields determine an activity of TTS and play a crucial role in
an evolution of angular momentum of these objects, therefore a question
about strength and topology of the magnetic field is one of the fundamental
questions of physics of young stars. Magnetic field was found in $\sim 10$
TTSs at the moment, and for some of them an information about its structure
has been retrieved -- see Donati \& Landstreet (2009) and references therein.

  RW Aur was included in the original list of T Tauri stars
compiled by Joy (1945). The year before Joy and van Biesbroeck (1944) have
discovered that the star (RW Aur A) has a companion (RW Aur B) at the
distance $\simeq 1.2^{\prime \prime}$. It was found later that both
components are so-called classical T Tauri stars (CTTS), whose activity is
due to accretion of matter from protoplanetary disc (Bertout, 1989). Recently
images of the discs were obtained for both components of the system in the
submillimeter region (Cabrit et al., 2006). Contribution of the secondary to
the overall optical flux does not exceed a few percent (Petrov et al.,
2001), therefore it can be assumed that features of optical spectra
described below belong to the primary.

   Optical spectra of RW Aur were investigated by many researches -- see
Petrov et al. (2001); Alencar et al. (2005); Petrov \& Kozack (2007) and
references therein. It was found that the spectrum has a multicomponent
structure which includes:

  1) a photospheric absorption K1-K4 spectrum superimposed with
emission continuum and narrow absorption lines with radial velocities close
to that of photospheric lines;

  2) absorption components in red wings of some lines that indicate gas
inflow onto the star with velocity up to 400 km\,s$^{-1}$;

  3) absorption components at blue wings of some lines, which indicate a gas
outflow from stellar vicinity with a velocity greater than 200
km\,s$^{-1}$. The outflowing gas is collimated into two jets with an
opposite orientation at large distances from the star (Hirth et al. 1994).
Lines of [O\,I] and [S\,II] are originated in these jets;

  4) emission lines of different elements which are (in a differnt degree)
superposition of a broad (FWHM$\sim 100-200$ km\,s$^{-1}$) and
narrow (FWHM $\sim 50$ km\,s$^{-1}$) components. The components are most
likely formed in different regions.

  Petrov et al. (2001) have found periodic variations of radial velocities
of photospheric lines with an amplitude about 6 km\,s$^{-1}$ and a period
$\simeq 2.77$ days. The same period was found for variations of radial
velocity $V_r^{NC}$ an equivalent width $EW_{NC}$ of narrow emission
component of He\,I 5876 line as well as an equivalent width EW$_{RA}$ of
an absorption feature in line's red wing. Velocities $V_r^{ph}$ and
$V_r^{NC}$ vary in antiphase and phase curves of $EW_{NC}$ and
$EW_{RA}$ are shifted relative to the curve $V_r^{ph}$ to one fourth of
the period. Similar behaviour of the narrow components and absorptions
at the red wing is also observed for some other lines. What about
broad components of emission lines it turned out that they vary with a
period of about 5.5 days, i.e. twice slower than $V_r^{ph}.$

  These features of RW Aur spectrum can be explained by a model of
non-axisymmetric magnetospheric accretion. Petrov et al. (2001) have
considered two possible reasons for a violation of an axisymmetry. The
first reason is an existence of low-mass companion with an orbital period
of 2.77 days. It was assumed that an interaction of the companion and the
disc results in a fall of disc matter towards RW Aur A in a form of
stream, which rotates together with the satellite around the central star.

  An alternative hypothesis assumes that RW Aur A is a single star with a
period of axial rotation $P\simeq 5.5$ days and a magnetic axis is tilted
relative to a rotational axis. If stellar magnetic field is not too much
differ from a dipole field then gas accreting from the disc forms two
streams, falling onto opposite hemispheres of the star (Romanova et al.,
2003). So-called hot spots arises at the footstep of the streams such as
both spots should be visible in the case of RW Aur A what can explain why
variations of photospheric and emission lines parameters occur
with a period twice smaller than rotational one.

  Magnetic field measurements allow to choose between these hypotheses
because in the case of a single star magnetic field in hotspots should
changes its sign to opposite after a half of the period.

  The only attempt to measure magnetic field in RW Aur was made by
Symington et al. (2005) who observed the star three times: December,5 2001,
December, 21 and 23 2002. But results of these measurements even averaged
over night have low precision ($0.46 \pm 0.24$ kG, $0.20 \pm 1.18$ kG,
$0.28 \pm 0.39$ kG correspondingly) and cannot be used to choose between the
models of a single and a binary star.

  New results of measurements of magnetic field longitudinal component in RW
Aur A are presented here. We succeded to confirm the model of two
hotspots with opposite magnetic field polarities and to found strong
magnetic field (up to $\simeq 0.8$ kG) in RW Aur A's outflow.

                             %%%%%%%%%%%%%%%%%%%

\section*{Observations and data reduction}

  The method we use to measure magnetic field is based on the fact that
so-called $\sigma$-components resulted in Zeeman splitting are polarized
circularly such as oppositely polarized components are located on different
sides of the central wavelength $\lambda_0$. If magnetic field in a
line formation region has a longitudinal component $B_z$ then the right-
and left-hand polarized components will be shifted relative to each other to
$$
\Delta\lambda_{rl} \simeq 2.3 \cdot 10^{-2} \,g\,
{\left( {\lambda_0 \over 5000} \right)}^2 \, B_z,
$$
where g is the Lande g-factor of the line under consideration,
$\Delta\lambda_{rl},$ $\lambda_0$ are in {\AA} and $B_z$ is in kG.
This relation allows to found the longitudinal magnetic field component
averaged over the line formation region by measuring $\Delta\lambda$ from
two spectra observed in the right- and left-hand polarized light.

  Our observations were carried out in January-February 2006, December 2007,
January 2008 and January 2009 with the Main Stellar Spectrograph (Panchuk,
2001) of 6-m telescope of Special Astrophysical Observatory
equipped with a polarization $\lambda/4$-plate and a double slicer
(Chountonov, 2004). 18 spectra of RW Aur have been observed. The
spectrograph slit width $0.^{\prime\prime}5$ provided a spectral
resolution of $R\sim15000$ in $5540-6600$ \AA\AA\, spectral band.

  Spectra were processed as follows (Chountonov et al., 2007). Dark current,
night sky emission and detector bias as well as cosmic ray traces were
removed in a standard way, using routines from the MIDAS software package. A
spectrum of a thorium-argon lamp was used for wavelength calibration. The
spectra are transformed into the stellar rest frame, i.e. the Earth radial
velocity and the average radial velocity of RW Aur ($+10.5$ km\,s$^{-1}$)
were subtracted. Hereafter all velocities of spectral features will be
specified in the stellar rest frame.

   Each observed spectrum covered a range $\Delta \lambda$ about 360 \AA\,.
The center of spectra was near He\,I 5876 or Na\,I D lines in 11 cases and
shifted to the red on a few hundred \AA\, in other 7 spectra. We will
refer to these spectra as "blue"\, and "red"\, respectively. The log of
observations is presented in Table \ref{tab-zhurnal}.

  To measure the difference between line positions in the spectra with
opposite polarizations we used a version of the crosscorrelation method
(Johnstone \& Penston, 1986) for the confidence level $\alpha=0.68$ what
corresponds to 1\,$\sigma$ error. To exclude systematic instrumental errors
our observations were organized as follows.

  Between exposures the superachromatic quarter-wave phase plate was
rotated in such a way that the right- and left-hand polarized spectra on the
CCD array changed places. Let $\Delta\lambda^{(1)}_{rl}$ be the difference
between the positions of a certain line in the spectra with opposite
polarizations measured at the initial position of the phase plate
and $\Delta\lambda^{(1)}_{rl}$ be the same quantity measured after
the plate's rotation. Then the difference
$$
\Delta\lambda_{rl}=\frac{\Delta\lambda^{(1)}_{rl}+\Delta\lambda^{(2)}_{rl}}{2}
$$
is free from systematic errors the main of which is spectrograph's slit tilt.
Thus two exposures of the star are required to measure the line shift.
Exposure time was about 20 minutes. In Table~\ref{tab-zhurnal} each pair
of expositions labeled as one observation, hence each estimation
of $B_z$ was carried out using 4 spectra.

  The signal-to-noise ratio from Table~\ref{tab-zhurnal} is estimated as
$\sigma_V^{-1}$, where $\sigma_V$ -- standard deviation calculated on Stokes
$V$-curve, from which a large-scale trend was subtracted.
\footnote{ Stokes $I$-curve is a half-sum of spectra with the left and right
polarization, i.e. $I(\lambda) =(F_\lambda^r + F_\lambda^l)/2,$ and
the $V$-curve is defined by the relation:
$V(\lambda) = (F_\lambda^r - F_\lambda^l)/I(\lambda).$
}
%

                 %%%%%%%%%%%%%%%%%%%%%%%%%%
\begin{table}
\caption{Log of observations.}\label{tab-zhurnal}
\begin{center}
\begin{tabular}{|c| c |c |c |c|}
\hline
 JD 245... & $\Delta \lambda$, {\AA} & SNR & $V_r^{ph},$ km\,s$^{-1}$
& $\sigma_{Vr},$ km\,s$^{-1}$ \\
\hline
3746.30 & 5575.7 - 5941.6 & 370 &  2.3 &  3.6 \\
3748.30 & 5575.2 - 5941.2 & 790 & 10.6 &  1.7 \\
3749.33 & 5575.1 - 5941.2 & 520 &  2.6 &  2.0 \\
3784.29 & 5741.2 - 6107.2 & 450 & 17.0 &  1.5 \\
3784.41 & 6166.6 - 6532.1 & 370 & 13.3 &  2.0 \\
3786.43 & 5743.7 - 6109.5 & 390 & 15.9 &  1.7 \\
3786.49 & 6169.0 - 6534.3 & 300 & 18.9 &  1.8 \\
4460.26 & 5585.2 - 5951.1 & 430 &  6.7 &  1.7 \\
4460.56 & 6074.2 - 6439.7 & 590 & 16.1 &  2.3 \\
4460.69 & 5599.0 - 5964.9 & 380 &   -  &   -  \\
4461.25 & 5599.1 - 5965.1 & 560 &  4.2 &  1.3 \\
4461.41 & 6075.4 - 6440.9 & 670 & 16.6 &  3.0 \\
4461.56 & 5630.6 - 5996.5 & 610 &  5.7 &  2.0 \\
4461.68 & 6070.9 - 6436.4 & 320 & 11.9 &  2.8 \\
4486.27 & 5737.4 - 6103.2 & 470 &  4.6 &  1.6 \\
4486.47 & 6218.3 - 6583.5 & 590 &  8.7 &  3.9 \\
4846.34 & 5796.0 - 6161.8 & 480 & 10.8 &  0.7 \\
4846.44 & 6247.2 - 6612.3 & 510 & 11.4 &  1.7 \\
\hline
\end{tabular}
\end{center}
\end{table}
                 %%%%%%%%%%%%%%%%%%%%%%%%%%

  We observed magnetic star 53 Cam several times to test our instrumental
set-up and data reduction process. For example January,20 2008 we found
$B_z=0.34\pm 0.11$ kG what is in a good agreement with the ephemeris
value $B_z=0.35$ kG (Hill et al., 1998).

  In some cases we subtracted a veiled photospheric spectrum from
observed one using the method described by Hartigan et al.
(1989). The template star was HD\,138716 (spectral type K1\,IV), a
spectrum of which was retrieved from UVES library (Bagnulo et al. 2003)
and corrected for radial velocity of $+50$ km s$-1$ derived by
comparison with theoretical spectrum from the VALD database (Kupka et
al., 1999). Lines in HD\,138716 spectrum were artificially
broadened by a convolution with Gaussian with $\sigma=15-20$ km\,s$^{-1}$.

%%%%%%%%%%%%%%%%%%%%%%%%%%%%%%%%%%%%%%%%%%%%%%%%%%%%%%%%%%%%%%%%%%%%%%%%%%%%%

\section*{Results}

   The following lines were used in our blue spectra to measured $B_z:$
photospheric lines (we correlated them together), He\,I 5876,
Na\,I D$_2$ 5890.0 and Na\,I D$_1$ 5895.9. Fe\,I 5659 line also was used
when it fell inside the wavelength band. Results of these measurements
are listed in Table\,\ref{blue-band}.

  Generally speaking He\,I 5876 line consists of six fine structure
components resulted from allowed transitions between $2p$ ${}^3P_{0,1,2}^o$
and $3d$ ${}^3D_{1,2,3}$ levels. The maximum wavelength difference between
components is 0.36 \AA\, or $\simeq 18$ km\,s$^{-1}$ in velocity scale. The
central wavelength was accepted to be equal to 5875.6 {\AA} derived by
averaging component's wavelengths with weights proportional to statistical
weights of their upper level. He\,I 5876 line's Lande factor $g$ was
accepted to be equal to 1.1.

  We supposed that mentioned above Fe\,I 5659 emission line is a blend of
three Fe\,I lines with $\lambda$ 5658.53 \AA, 5658.66 \AA\, and 5658.82 \AA.
We consider this blend as one line with Lande factor $g=1.0$.

      %%%%%%%%%%%%%%%%%%%%%%%%%%%%%%%%%
\begin{table}
\caption{Results of $B_z$ measurements from blue spectra. }\label{blue-band}
 \begin{center}
  \begin{tabular}{|c | c | c| c c | c c | c c | c l|}
\hline
& & & \multicolumn{2}{|c|}{He\,I 5876} & \multicolumn{2}{|c|}{Na\,I D}
& \multicolumn{2}{|c|}{Fe\,I 5659} & \multicolumn{2}{|l|}{The Photosphere} \\
N & JD 245... & $\varphi$ &\multicolumn{1}{|c}{$B_z$} & $\sigma_B$
& \multicolumn{1}{|c}{$B_z$} & $\sigma_B$
& \multicolumn{1}{|c}{$B_z$} & $\sigma_B$
& \multicolumn{1}{|c}{$B_z$} & $\sigma_B$ \\
\hline
 1 & 3746.30 & 0    &$-0.43$ & 0.25 & $+0.08$ & 0.08 & $+0.17$ & 0.30 & $+0.23$ & 0.10 \\
 2 & 3748.30 & 0.36 &$-1.47$ & 0.15 & $-0.18$ & 0.05 & $-0.15$ & 0.30 & $+0.07$ & 0.20 \\
 3 & 3749.33 & 0.54 &$-1.01$ & 0.20 & $-0.81$ & 0.08 & $-0.10$ & 0.36 & $+0.11$ & 0.30 \\
 4 & 3784.29 & 0.81 &$+1.10$ & 0.15 & $-0.03$ & 0.08 & $     $ &	& $-0.83$ & 0.30 \\
 5 & 3786.43 & 0.20 &$-0.11$ & 0.35 & $+0.13$ & 0.08 & $     $ &	& $-0.04$ & 0.20 \\
\hline
 6 & 4460.26 & 0.04 &$-0.82$ & 0.25 & $+0.05$ & 0.08 & $+0.36$ & 0.30 & $+0.26$ & 0.20 \\
 7 & 4460.69 & 0.12 &$-0.98$ & 0.25 & $-0.02$ & 0.08 & $+0.64$ & 0.42 & $+0.03$ & 0.30 \\
 8 & 4461.25 & 0.22 &$-1.04$ & 0.15 & $+0.02$ & 0.08 & $+0.47$ & 0.36 & $+0.12$ & 0.20 \\
 9 & 4461.56 & 0.27 &$-1.11$ & 0.15 & $+0.09$ & 0.05 & $+0.11$ & 0.48 & $+0.12$ & 0.20 \\
10 & 4486.27 & 0.71 &$+0.50$ & 0.20 & $-0.11$ & 0.08 & $     $ &	& $-0.02$ & 0.20 \\
\hline
11 & 4846.34 & 0.28 &$-1.19$ & 0.15 & $+0.03$ & 0.08 & $     $ &	& $+0.10$ & 0.10 \\
\hline
\multicolumn{11}{l}{Magnetic field strength $B_z$ and its uncertainty
$\sigma_B$ are expressed in kG.} \\
\multicolumn{11}{l}{$\varphi$ -- phase of axial rotation for $P=5.576^d$ period.
} \\
  \end{tabular}
 \end{center}
\end{table}

			%%%%%%%%%%%%%%%%%%%%%%%%%%%%%%

  Absorption photospheric lines, [O\,I] 6300.3 line, lines of Si\,II
$6347.1+6371.4$ doublet and Fe\,II 6432.7 line always fell into the red
spectra in contrast to Fe\,II 6456.4 and Fe\,II 6516.1 lines.
To measure $B_z$ in the photosphere we correlate all appropriate lines
together as well as in the case of field's measurement in the formation
region of Si\,II or Fe\,II lines. Lande factors of Fe\,II and photospheric
lines were adopted from the VALD database (Kupka et al., 1999) and assumed
to be equal to 1 for [O\,I] 6300.3 and Si\,II lines. In the case of blended
photospheric lines $g$-factors were averaged with weights of a central
depth. The results of the field measurements for the red spectra are
presented in Table~\ref{red-band}.

  $H_{\alpha}$ line twice (JD 2454486.47 and 2454846.44) fell into the
spectral region and we found $B_z=+130 \pm 100$ G and $+40 \pm 100$ G
in its formation region correspondingly assuming $g=1.$

			%%%%%%%%%%%%%%%%%%%%%%%%%%%%%%
\begin{table}
\caption{Results of $B_z$ measurements from red spectra.} \label{red-band}
\begin{center}
\begin{tabular}{|c | c c | c c | c c | c l|}
\hline
 & \multicolumn{2}{|c|}{[O\,I] 6300}
 & \multicolumn{2}{|c|}{Si\,II}
 & \multicolumn{2}{|c|}{Fe\,II}
 & \multicolumn{2}{l|}{The photosphere} \\
JD 245...  &
\multicolumn{1}{|c}{$B_z$} & $\sigma_B$ &
\multicolumn{1}{|c}{$B_z$} & $\sigma_B$ &
\multicolumn{1}{|c}{$B_z$} & $\sigma_B$ &
\multicolumn{1}{c}{$B_z$} & $\sigma_B$ \\
\hline
3784.41 & $+0.32$ & 0.24 & $-0.03$ & 0.20 & $-0.09$ & 0.20 & $-0.22$ & 0.20 \\
3786.49 & $-0.12$ & 0.24 & $-0.58$ & 0.60 & $-0.11$ & 0.20 & $+0.18$ & 0.20 \\
4460.56 & $+0.06$ & 0.29 & $+0.07$ & 0.40 & $	  $ &	   & $+0.05$ & 0.10 \\
4461.41 & $-0.08$ & 0.19 & $-0.89$ & 0.40 & $	  $ &	   & $+0.17$ & 0.10 \\
4461.68 & $-0.09$ & 0.19 & $-0.35$ & 0.60 & $	  $ &	   & $+0.15$ & 0.20 \\
4486.47 & $+0.20$ & 0.14 & $+0.08$ & 0.50 & $+0.02$ & 0.10 & $-0.04$ & 0.10 \\
4846.44 & $+0.06$ & 0.14 & $-1.17$ & 0.50 & $-0.11$ & 0.10 & $+0.07$ & 0.10 \\
\hline
\end{tabular}
\end{center}
\end{table}

			%%%%%%%%%%%%%%%%%%%%%%%%%%%%%%

  Magnetic field was not detected at $3\sigma$-level in photospheric lines
and at $2\sigma$-level in formation regions of H$_\alpha,$ Fe\,I,
Fe\,II, Si\,II and [O\,I] 6300 lines. $B_z$ upper limits
in the photosphere and in the [O\,I] 6300 line formation region (after
averaging over all measurements at $3\sigma$-level) are 180 G and 230 G
correspondingly.

  Fe\,I and Fe\,II emission lines have similar profiles and consist solely
of the broad component. Therefore using these lines we measure the field
in a region where the broad component is formed. As far as shapes and
equivalent widths of the broad components are variable (Petrov et al.,
2001) there is no sence to average the field over time. Depending on the night,
upper limits (at $3\sigma$-level) for Fe\,II lines lie between 300 and 600 G
-- see Table~\ref{red-band}.

\begin{figure}[h!]
 \begin{center}
 \resizebox{14.0cm}{!}{\includegraphics{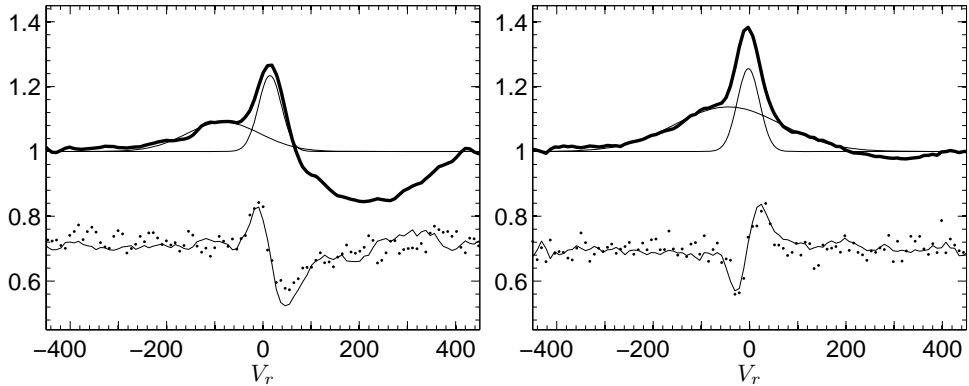} }
 \caption{ Stokes $I$ and $V$ profiles of He\,I 5876 line for
JD 2453748.30 (left panel) and JD 2453784.29 (right panel).
Solid bold line represents $I$-profile normalized to the continuum level.
Solid thin lines are Gaussians which correspond to a narrow and a broad
components. Dots represents observed Stokes $V$-profile on which model
$V$-profile was overplotted by a solid thin line. Both $V$-profiles were
multiplied by 10 times and shifted to 0.7 for readability. An abscissa is
the velocity in km\,s$^{-1}$ from the line center $\lambda_0=5875.6$ \AA.
}
  \label{He-I-V-profs}
  \end{center}
\end{figure}

  Magnetic field was detected in He\,I 5876 line formation region.
The field has a different sign in different moments -- see
Fig.~\ref{He-I-V-profs} on which spectrum of a veiled photosphere
was subtracted from Stokes $I$-profile.

			%%%%%%%%%%%%%%%%%%%%%
			%%%%%%%%%%%%%%%%%%%%%

  For illustrative purposes we computed model $V$-profiles. After the field
and corresponding shift $\Delta \lambda_{rl}$ between right- and left
polarized profiles had been derived, we computed the difference between two
$I$-profiles, the first of them was shifted by $\Delta \lambda_{rl}/2$ to
the left and the second was shifted by the same amount to the right. Having
divided the difference by $I(\lambda)$, we simulated a model $V$-profile. It
can be seen that in an emission part of the $I$-profile the model
$V$-profile well approximates observed one and clearly indicates that
the field in observations presented on the left and right panels of
Fig.\,\ref{He-I-V-profs} has the opposite sign.

\begin{figure}[h!]
 \begin{center}
 \resizebox{14.0cm}{!}{\includegraphics{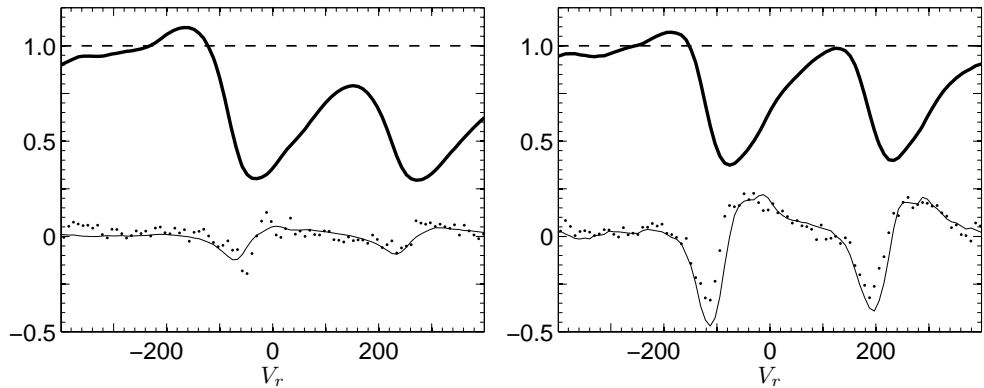} }
 \caption{ The same as in Fig. \ref{He-I-V-profs}, but for Na\,I D doublet
lines. The left panel corresponds to the moment JD 2453748.30 and the right
panel to JD 2453749.33. In contrast to the previous figure observational
and model $V$-curves, although were multiplied by 10 times, but were not
shifted along ordinate axis. In addition the spectrum of a veiled photosphere
was not removed on this figure. Zero point of abscissa axis corresponds to
the center of D$_2$ line $(\lambda=5889.95$~\AA).
}

  \label{Na-I-V-profs}
  \end{center}
\end{figure}

  Also we have detected twice (JD 2453748.30 and 2453749.33) at $>3\sigma$-level
$B_z$ in Na\,I D line formation region -- see Table~\ref{red-band}. Lande
factor for both lines of the doublet was assumed to be equal $1.33$
(Frish, 2010). The Stokes $I$ and $V$ profiles for Na\,I D lines at the
mentioned moments are shown on Fig.~\ref{Na-I-V-profs}. Model
$V$-profile for Na\,I D lines was computed in the same manner as for He\,I
5876 \AA\, line.

			%%%%%%%%%%%%%%%%%%%%%
			%%%%%%%%%%%%%%%%%%%%%

  In the conclusion of this section note the following. It can be found from
table 1 of Petrov et al. (2001) paper that RW Aur's radial velocity of
photospheric lines averaged over all obsrvations with $\sigma^{-2}$ weight
is equal to $+16.0$ km\,s$^{-1}$. The same quantity derived from our
observations is equal to $+10.5$ km\,s$^{-1}$. Possible reasons of the
discrepancy will be discussed in the Conclusion.

			%%%%%%%%%%%%%%%%%%%%%%%%%%%%%

\section*{Magnetic field in the hotspots}

   As was noted in the Introduction some lines in RW Aur's spectrum
have redshifted absorption feature extended up to the velocity
$\simeq 400$ km\,s$^{-1}$ (see e.g. the left panel of
Fig.\,\ref{He-I-V-profs}) indicating matter infall with the same velocity.
Accreted gas should be deccelerated near stellar surface in the shock and
abruptly heated up at the shock front to $T\sim 10^6\,K.$ Post shock zone is
the region were thermal and kinetic energy of accreted matter is radiated
away in X-ray and far UV bands so gas cools down, deccelerates and finally
gradually settles to the stellar surface. The source of an emission
continuum are regions at the footstep of accretion streams, which absorb the
hard radiation of the shock and re-emit it mainly in the region from 0.3 to
1 $\mu.$ These regions are used to call hotspots although their temperature
is $<9000$ K (Lamzin, 1995; Calvet \& Gullbring, 1998).

  It was found about 20 years ago that many emission lines in CTTS's spectra
are superposition of so-called narrow and broad components which probably
formed in regions with different physical conditions -- see e.g. Batalha et
al. (1996) and references therein. In the case of RW Aur lines of ionized
metals and He\,I have two-component structure, He\,II 4686 has no broad
component and vice versa hydrogen lines and other neutrals have no
significant narrow component (Petrov et al., 2001).

  Fig.\,\ref{He-I-V-profs}) demonstrates how emission part of He\,I 5876
line's $I$-profile can be decomposed onto two Gaussians, which correspond to
narrow and broad components. As was mentioned in the Introduction
Petrov et al. (2001) have found that radial velocity of the narrow component
$V_r^{NC}$ varied with a period of about 2.77 days around a mean velocity
$\simeq +10$ km\,s$^{-1}$ with an amplitude $\simeq 15$ km\,s$^{-1}$.

  Petrov et al. (2001) suggested that narrow components are formed in
the hotspot(s) i.e. at the stellar surface. But it would be more precisely
to say that they are formed {\it above} the photosphere of the hotspot, i.e.
in the post shock cooling zone where helium atoms and first ions of metals
appear due to recombination from higher ionization states. This remark
helps to explain in natural way why narrow components have small positive
velocity relative to the photosphere. It becames also clear why radial
velocity of He\,II 4686 line is greater than $V_r^{NC}$ of He\,I 5876 line
(Petrov et al., 2001): He$^+$ ions appears behind the shock front before
helium atoms and thus have greater velocities.

  However in the case of RW Aur the accretion shock is placed very close to
the stellar surface. Indeed an existence of a veiling continuum with a flux at
$\lambda =5500$ \AA\, exceeding the photospheric one by 2-3 times (Petrov et
al., 2001) suggests that pre-shock gas density is greater than
$10^{11}$ sm$^{-3}.$ Hence, a distance between the shock front and the
stellar photosphere is much smaller than the stellar radius (Lamzin, 1995)
what means that narrow components are formed practically at the stellar
surface.

  It can be seen from the comparison of observed $I$-profile and
modeled $V$-profile on Fig. \ref{He-I-V-profs} that the difference
$F_\lambda^r - F_\lambda^l$ is maximal where $I$-profile has a steep
gradient. Therefore the shape of an {\it observed} $V$-profile of He\,I 5876
does not constrain a region of the strong $B_z$ by only a formation region
of the narrow component. We found that if to assume {\it a priori} (i.e.
before cross corellation procedure) that $B_z=0$ in the broad component then
then field averaged over the narrow component increases by no more than
1\,$\sigma_B$.

  There is no doubt that the field changes its polarity. In the frame of
magnetospheric accretion model this fact can be naturally explained by a
change of an orientation of the hotspot relative to the Earth due to stellar
axial rotation. It means that the variations of the field should be
periodical with the period of the stellar axial rotation.

\begin{figure}[h!]
 \begin{center}
 \resizebox{12.0cm}{!}{\includegraphics{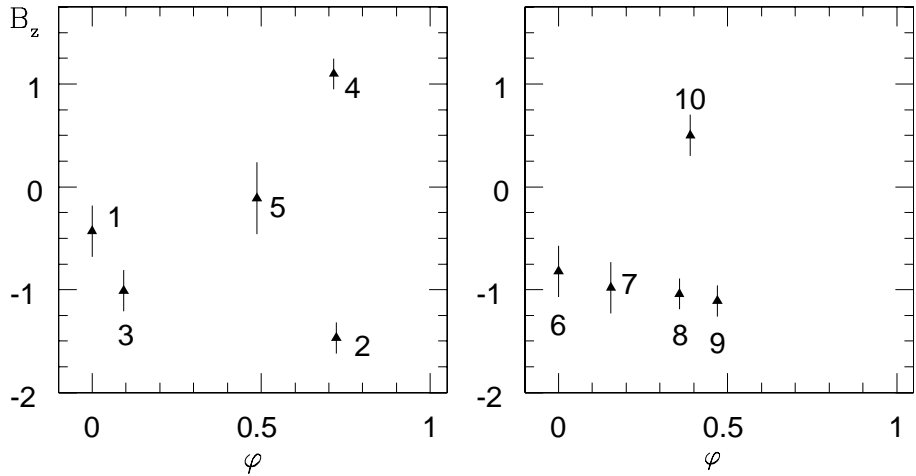} }
 \caption{ A phase curve of variations of $B_z$ (kG) in a region of the
He\,I 5876 line formation for $P=2.77$ days. Numeration of the dots on the
figure corresponds to the numeration of spectra in Table~\ref{blue-band}.
An ordinate is the field in kG. $\varphi=0$ phase corresponds to the dot
\# 1 on a right panel and to the dot \# 6 on a left panel.}
 \label{2.77-period-curve}
 \end{center}

\end{figure}

  On Fig.\,\ref{2.77-period-curve} phase curves of observed $B_z$ derived
over the He\,I 5876 line are shown for the period of 2.77 days. Our
observations cover a time span of a few years and during the time a geometry
and strength of the field could change: Donati et al. (2008) have discovered
such changes for CTTS BP Tau by a comparison of stellar magnetogramms
obtained with a time span of 10 month.

   To minimize the effect of a physical variability of the field, we
considered separately two sets of observations from Table~\ref{blue-band},
which contain 5 observations (in Table the sets are separated by a
horizontal line). The first set (left panel) contains observations, which
were carried out during 40 days and the second set (right panel) -- the next
five observations covered a time span of 26 days. A global changing of a
magnetic structure over times less than 40 days seems unlikely, therefore
the observational variations of the magnetic field during each set is caused
by the axial rotation. What about the last observation, it was carried out
360 days after tenth observation and does not shown on
Fig.\,\ref{2.77-period-curve}.

				%%%%%%%%%%%%%%%%%%%%%
			%%%%%%%%%%%%%%%%%%%%%

   The figure suggests that in the first set $B_z$ changes from $+1.1 \pm
0.15$ kG to $-1.47 \pm 0.15$ kG during $\simeq 15$ minutes (the time, which
correspond to a phase difference of the dots 4 and 2). In the second set
nearly the same changing of the field strength occurred during a time span
of $<50$ minutes (the dots 8 and 10), and the field recovered to an initial
level after $<3$ hours (the dot 9). Such kind of variations cannot be
explained by only an orbital motion of the star with $P\simeq 2.77$ due to
an existence of a close companion. We found that the same statement is valid
for any period in the range from 2.5 to 2.9 days that includes all possible
values of a "short"\, $(\sim 2.7^d)$ period for RW Aur A (Petrov et al., 2001).

\begin{figure}[h!]
 \begin{center}
  \resizebox{12.0cm}{!}{\includegraphics{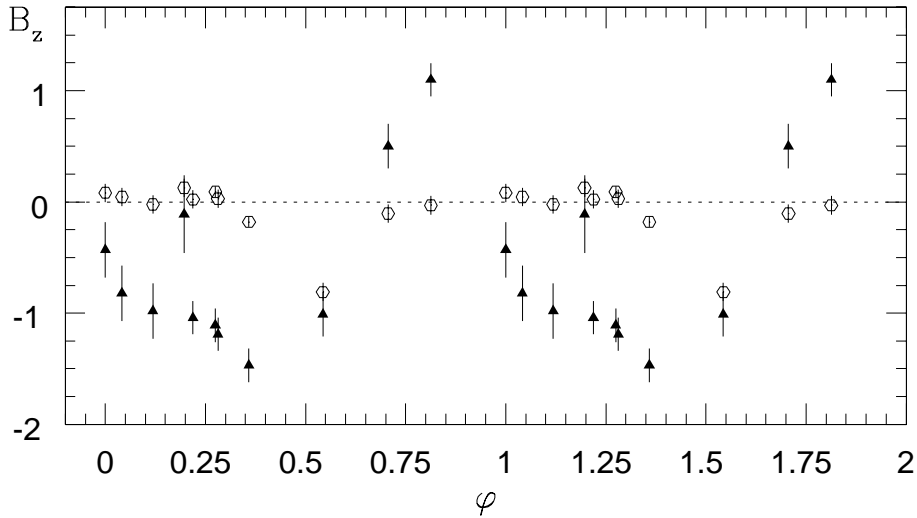} }
  \caption{ Phase curve of variations of $B_z$ (kG) with $P=5.576^d$
for He\,I 5876 line (triangles) and for Na D doublet lines (open circles).
}
  \label{5.58-period-curve}
  \end{center}
\end{figure}

   One the other hand, we have found that for a number of values $P$ in the
range from 5.5 to 5.7 days a smooth dependence $B_z=B_z(\varphi)$ could be
obtained even without a phase shift between the sets. An example of such
curve is shown on Fig.~\ref{5.58-period-curve} for $P=5.576^d$ but
similar smooth curve turns out for $P=5.6659^d.$ In other words for $P\sim 5.6$
days observed variability of $B_z$ can be explained as the result of
rotational modulation only, i.e. assuming that geometry and field strength
in the accretion zone did not change significantly during three
years of the observations.

				%%%%%%%%%%%%%%%%%%%%%
				%%%%%%%%%%%%%%%%%%%%%

   Having only 11 measurements of $B_z$ we cannot derive a certain period of
the axial rotation of RW Aur A as well as make a conclusion concerning a
variability of a geometry and/or physical conditions in the accretion zone
from January 2006 to January 2009. However, we can certainly argue that our
results contradict to the rotational period $\simeq 2.77^d,$ but well consist
with $P\simeq5.6^d,$ which corresponds to the model of two hotspots with
opposite polarities of the field and located in opposite hemispheres.

   In the frame of the model observed peculiarities of RW Aur A
spectrum can be explained in the following way (Petrov et al., 2001).
Temperature increases with height above hotspot's photosphere what results
in appearence of emission lines. The strongest lines appear in the spectrum
as the narrow components of helium and ionized metals lines, but more faint
lines blend absorption lines of the photosphere to some extent. The hotspots
move relative to the Earth due to stellar rotation. Due to this reason
radial velocities of narrow components $V_r^{NC}$ vary as well as a center
of gravity of the photospheric lines what looks like a variation of their
radial velocity $V_r^{ph}.$ This effect explains why $V_r^{NC}$ and
$V_r^{ph}$ vary in the antiphase.

  It can be explained in the same way why a phase curve of He\,I 5876
line's narrow component equivalent width $EW_{NC}$ is shifted by one fourth
of the rotational period relative to $V_r^{NC}(\varphi)$ curve. Indeed
when the center of each spot crosses the plane formed by the Earth and
stellar rotational axis (the central meridian), $V_r^{NC} \simeq 0$ and
observed area of the spot and therefore $EW_{NC}$ have maximum values. After
1/4 of the period the spot center is close to a limb and modulus of
$V_r^{NC}$ has a maximum but $EW_{NC}$ has a minimum.

   Our data are insufficient to reconstruct a magnetic map of the star,
so we will give below only semi-qualitative constraints on a geometry and
parameters of accretion zones.

\begin{figure}[h!]
 \begin{center}
  \resizebox{12.0cm}{!}{\includegraphics{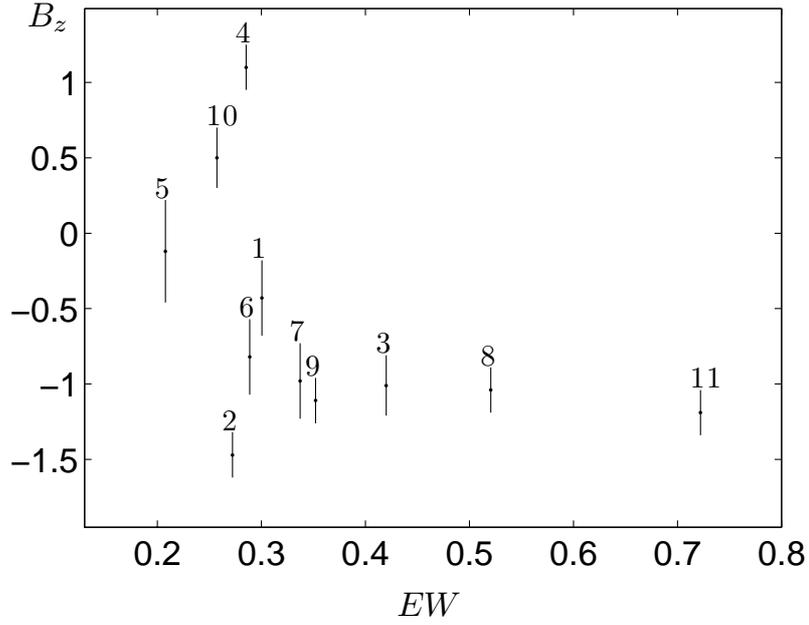} }
  \caption{ A dependence of $B_z$ modulus (kG) on
equivalent width of the narrow component of the He\,I 5876 line (\AA).}
  \label{EW-He5876-vs-B}
  \end{center}
\end{figure}

 A relation between $B_z$ measured over the He\,I 5876 line and an
equivalent width of the narrow component of the line is shown on
Fig.\,\ref{EW-He5876-vs-B}. Bear in mind written above it follows from the
figure that maximal values of $\mid B_z\mid$ is reached when the hotspots
cross the central meridian and are close to zero when hotspots are
near stellar limb. One can conclude therefore that magnetic field lines in
the spots make large angle relative to stellar surface, what means that
toroidal component of the field in the spots is much less than poloidal one.

			%%%%%%%%%%%%%%%%%%%%%
			%%%%%%%%%%%%%%%%%%%%%

\begin{figure}[h!]
 \begin{center}
  \resizebox{10.0cm}{!}{\includegraphics{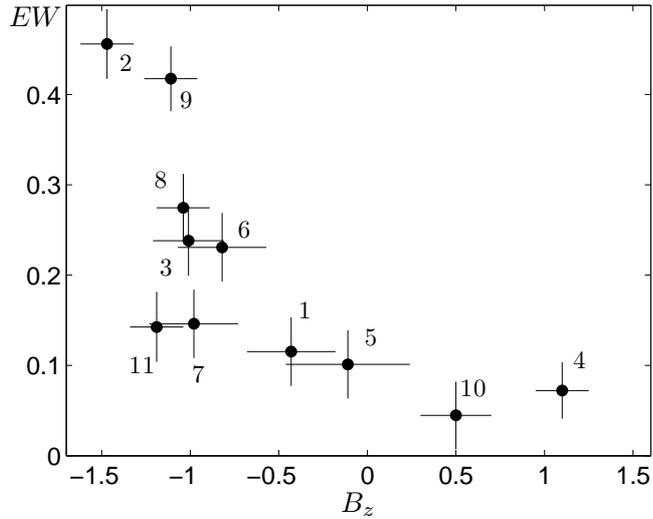} }
  \caption{A dependence of equivalent width of absorption feature (\AA) of the
  He\,I 5876 line derived in the range from $+200$ to $+400$ km\,s$^{-1}$
  on the field $B_z$ (kG).}
  \label{EW-He-abs-vs-B}
  \end{center}
\end{figure}

  It can be seen from Fig.\,\ref{He-I-V-profs} that when $B_z$ has large
negative value (the left panel) the helium line demonstrates deep
redshifted absorption feature which has velocity extention up to $+400$
km\,s$^{-1}$ and vice versa in the case of large positive field
(the right panel) we observe shallow absorption. It is a particular case of
the general trend: when the field increases from negative to positive
values, the equivalent width of red absorption decreases. It can be seen
more clear if to use equivalent width of the far redshifted part of the
profile and additionaly subtract veiled photosphere -- see
Fig.\,\ref{EW-He-abs-vs-B}. Note that the main uncertainty in $EW_{RA}$ is
due to uncertainty in veiling determination.

			%%%%%%%%%%%%%%%%%%%%%
			%%%%%%%%%%%%%%%%%%%%%

  Because negative values of $B_z$ appear more frequently than positive
$B_z$, we conclude that a spot with $B_z<0$ is located in a hemisphere above
a midplane of an accretion disc relative to the Earth (the "upper"{} spot)
and a spot with $B_z >0$ is below the disc (the "low"{} spot). One can
conclude that longitudes of the upper and low spots differ by $\simeq 180^o$
because variations of photospheric lines radial velocities are well
convolved with period $\simeq 2.77^d\simeq P/2.$ The lower spot can be
observed from the Earth if its stellar latitude $\theta<i,$ where $i\simeq
65^o$ is the inclination of the stellar rotational axis to the line of sight
(see the end of this section).

  Maximal (absolute) values of field $\left|B_z\right|$ we observed in
the upper and low spots differ from each other 1.5-2 times. If spots have
nearly identical parameters then the same ratio must take place for
averaged
\footnote{Precisely, a mean weighted with He\,I 5876 line flux.}
cosines of angles between the field lines and the line of sight for moments
when the spots cross the central meridian. It follows from the modelling
we carried out (assuming the field of the star is dipol and
intensity in the He\,I 5876 line is nearly isotropic) that such cosine ratio
could be obtained if spots located at latitudes $30^o<\theta <50^o.$

  Red wing absorption of the helium line is formed in accreted gas of
pre-shock zone. It can be easily seen that when the upper spot is visible
from the Earth accretion stream is projected onto larger area of stellar
surface than in the case when we observe low spot. We suppose that just due
to this reason $EW_{RA}$ is much less when upper spot crosses the
central meridian than in the case when low spot passes throug the same
meridian.

  As was noted before in the case of RW Aur A accretion shock front is
located very close to stellar surface. Therefore pre-shock zone is always
projected onto underlying hotspot. Apparently it is the reason why $EW_{RA}$
has never reached zero. It is interesting to note that absorption feature in
the red wing of the He\,I 5876, Na\,I 5898 as well as O\,I 7773 lines (Petrov
et al., 2001) extends up to +400 km\,s$^{-1}$ {\it in all} spectra that we
know. In other words at any position of the spots relative to the Earth we
observe the constant red cutoff velocity of the red wing. It can be
explained by assuming that $V_m=400$ is the largest velocity of the
accreting gas and the field lines come out of the spots in a broad interval
of angles.

  If the mass of RW Aur A $M_* \simeq 1.3\,M_\odot$ (White \& Ghez, 2001)
then corotation radius for rotational period $P\simeq 5.6$ is:
$$
R_c = {\left( {GM_* P^2\over 4\pi^2} \right)}^{1/3}
\simeq 14.5\,R_\odot \simeq 7.6 \,R_*,
$$
where $R_*\simeq 1.9\,R_\odot$ is the stellar radius (see the end of
this section). As far as matter can not freez into magnetic field lines
at distance $r>R_c$ free-fall velocity at the
stellar surface is smaller than $V_u$ value defined by equation:
$$
{GM_* \over R_*} - {GM_* \over R_c} = {V_u^2 \over 2},
$$
what means that $V_u \simeq 480$ km\,s$^{-1}$.

  If to replace $V_u$ by $V_m$ then it follows from the same equation that
gas should start to fall along magnetic field lines from distance
$R_f\simeq 2.6\,R_* \simeq 4.9\,R_\odot$ to reach stellar surface with
velocity 400 km\,s$^{-1}$.

    Let us discuss in brief the question about the value of angle
between rotational axis of RW Aur A and the line of sight. Hirth et al.
(1994) have discovered a bipolar jet of RW Aur. Studying temporal variations
of the jet structure L{\'o}pez-Mart{\'i}n et al. (2003) have found that the
angle between the jet and the line of sight is $i=46^o \pm 3^o.$ Cabrit et
al. (2006) have found discs around both components of RW Aur and concluded
that the jet is oriented along circumprimary rotational axis of
RW Aur A disk, such as $i$ found by L{\'o}pez-Mart{\'i}n is a
lower limit and in fact $45^o<i<60^o.$ If the rotational axes of the disc
and the star are aligned, then the inclination of RW Aur A also must be
bounded between $45^o$ and $60^o.$ Similar estimate was derived by
Alencar et al. (2005) in modeling hydrogen and Na D line profiles in the frame
of disc wind model: $55^o < i < 65^o.$

  Petrov \& Kozack (2007) have argued that irregular light variability
of RW Aur A is mostly associated with eclipses by dust clouds.
Taking into account existed estimate $i\simeq 45^o$ of
L{\'o}pez-Mart{\'i}n et al. (2003) Petrov \& Kozack supposed that the dust
clouds are formed in the disc wind, because the star could not be occulted
by the disc itself at such inclination. However as far as $i=45^o$ is turned
out to be a lower limit one can suppose that the dust clouds are inhomogeneities
in the disc and then at least internal regions of the disc shoul be inclined to
the line of sight by an angle significantly greater than $45^o.$

  Note finally that observed value $v_e \sin i = 17.2 \pm 1.5$ km\,s$^{-1}$
(Hartmann \& Stauffer, 1989). If RW Aur A's rotational period is $P\simeq
5.6$ days then $R_* \sin i \simeq 1.9 \pm 0.2$ $R_\odot$ what can be
consistent with the estimate of the stellar radius $R_*=1.7 \pm 0.3$
$R_\odot$ (White \& Ghez, 2001) within uncertainties only at large $i.$
Taking in to account all these arguments we regard as reasonable
estimates $i\approx 65^o,$ $R_*\simeq 1.9$ $R_\odot.$

			%%%%%%%%%%%%%%%%%%%%%%%%%%%

\section*{A magnetic field at an outer boundary of a magnetosphere and in a wind}

  As follows from Tabl.\,\ref{blue-band} and \ref{red-band} $B_z$-values
did not differ always from zero within errors of measurements in formation
region of iron lines, which have only broad component. The example of JD
2453784 night (when we found $B_z=+1.10 \pm 0.15$ kG in He\,I 5876 line
formation region but $B_z=-0.09 \pm 0.20$ kG over Fe\,II lines) clearly
demonstrates that magnetic field strength is different in formation regions
of narrow and broad components of emission lines. Another example is JD
2454846 night when $B_Z$ in the broad component was $-0.11\pm 0.10$ kG but
$-1.19\pm 0.15$ kG in He\,I narrow component. Thus our data confirm the
viewpoint according to which narrow and broad components are formed in
different regions.

  G\'omez de Castro \& Verdugo (2003) argued that broad components are
formed at outer boundary of stellar magnetosphere within an internal
ring-like region of the disk. A half-width of the broad components in RW Aur
spectra, except for H$_\alpha$-line, is $\simeq 200$ km\,s$^{-1}$
at continuum level $\simeq200$ km\,s$^{-1}$ what consists with a
projection of orbital (keplerian) velocity to the line of sight
at distance $R_f\simeq 2.6\,R_*$ from the star:
$$
V_K = \sqrt{{ GM\over R_f }}\, \sin i.
$$
Remind that $R_f$ is the distance from which free-falling gas
reaches velocity about 400 km\,s$^{-1}$ at the surface of the star.

  Another nontrivial result of our observations is detection of magnetic
field in the formation region of Na\,I D lines in two spectra out of 11.
The detection is out of doubts in these two cases: firstly $B_z$
differs from zero at more than $3\,\sigma$ (see Table~\ref{blue-band}) and
secondly $V$-curves have similar shapes for both lines of the doublet -- see
Fig.~\ref{Na-I-V-profs}. It is very important to note that field strength up
to $\simeq 0.8$ kG was found in the region where line's profile from $\simeq
-180$ to 0 km\,s$^{-1}$ is formed, i.e. in the outflow.

  Existence of sodium atoms in the wind indicates that gas temperature
in the outflow is relatively low $(T<10^4$ K) while gas moves with
velocities up to $V_w \sim 200$ km\,s$^{-1}$ what follows from an extension
of Na\,I D$_2$ line's blue wing. Thus $V_w$ is more than one order of
magnitude greater than local sound velocity, what means that the outflow is
driven by nonthermal, probably magneto-rotational mechanism, and the outflow
is launched from the disc rather than from stellar surface (Cabrit, 2007;
Matt \& Pudritz, 2007).

  According to Table~\ref{blue-band} magnetic field strength variations
occurred in the following way during the first three nights of our
observations. At the first night $B_z$ in the spots and in the wind was
zero at $3\,\sigma$-level. Two days later $B_z$ has reached its maximum
value $-1.45 \pm 0.15$ kG in the upper spot and was equal to $-0.18 \pm
0.05$ kG in the wind. At the next night $B_z$ has decreased to $-1.01 \pm
0.20$ kG in the upper spot but increased up to $-0.81\pm 0.08$ kG in the wind.

  Bear in mind discussion in the previous Section this sequence of
events can be interpreted in the following way. At the first night the upper
spot was located near stellar limb and then $B_z$ in the wind was zero.
Two days later when the spot has passed through the central meridian
wind's field became noticeble and reached its maximal value one day
later. Note that this conclusion (sequence of events) does not depend on
precise (unknown) value of stellar rotation period $P.$ But if
$P$ indeed equal to $5.576^d$ then it follows from phase curve shown on
Fig.\ref{5.58-period-curve} that the longitude component of wind's magnetic
field has detectable strength only at certain orientation of the
system relative to the observer, reaching its maximum value at approximately
$1/4$ of rotation period after passage of upper spot throug the central
meridian. Of course it is true only if the discovered field is related with
(quasi)stationary gas outflow rather than a fortuitous gas ejection.

  To explain observed profiles of H$_\beta,$ H$_\gamma$ and Na\,I D
lines that are originated in RW Aur outflow Alencar et al., (2005)
suggested phenomenological model of the disc wind which is launched from
narrow $(\Delta r/r \ll 1)$ region near inner boundary of accretion disc.
As far as the aim of these authors was to reproduce time-averaged line profiles
they considered axisymmetric outflow and achieved reasonable agreement with
observations.

  Similar picture of the outflow has been reproduced by Romanova et al.
(2009) in numerical MHD simulations. It turned out that in an axisymmetric
case, when stellar rotational axis coincides with magnetic dipole axis,
the outflow is launched from narrow region at the boundary between the disc
and stellar magnetosphere and forms thin-walled cone with a cone
half-angle $\simeq 30^o -40^o$ (so called conical wind). Test
simulations for non-axisymmetric case, when the angle between rotational and
magnetic axes was $30^o,$ have shown that the outflow in the cone occurred
as before from the internal boundary of the disc and with the same cone
angle but with azimuthal inhomogeneities of gas parameters. It can be said
that in the wind (above and below the disc) a spiral stream is formed.
Inside the stream gas density and field strength are higher than in the rest
part of the outflow.

  This picture is in a qualitative agreement with observations of RW Aur:
firstly, blueshifted absorption Na D$_2$ line is present permanently
(Alencar et al., 2005), secondly, a non-zero field $B_z$ was observed
only when the upper spot was near the central meridian.

 At the first glance it seems that observed value $B_z\simeq 0.8$ kG is
too strong for a wind because for the magnetic field strength about 2 kG
at the stellar surface the field strength at $r\simeq 2.6$ $R_*$ should be
$\sim 0.1$ kG. However in the model of conical wind the disc compresses stellar
magnetosphere what should increase the field strengs at the base of the stream.

		%%%%%%%%%%%%%%%%%%%%%%%%%%%%%%%%%%%%%%%

\section*{Conclusion}

  We obtained 18 measurements of RW Aur A average magnetic field's
longitudinal component between 2006 and 2009 yrs. Our data
scaterred in time and can not be used for Zeeman imaging but a number of
nontrivial conclusions were obtained.

  First of all one can definitely state that results of our observations
contradict to hypothesis that explains periodical variations of RW Aur A
photospheric line's radial velocity as the result of orbital motion with a
period of $\simeq 2.77^d.$ At the same time our data consist with Petrov
et.al (2001) model of two hotspots with opposite magnetic field polarities
that located in opposite hemispheres, such as stellar axial rotational
period $P\simeq 5.6^d.$

  Our data are insufficient to derive precise period but we found that for
$P=5.576^d$ and $P=5.6659^d$ $B_z$-phase curves have the following
features:

 1) maximum and minimum values of $B_z$ in the spots are shifted in phase
by $\simeq 0.5$ what is expected in the case of two spots with longitude
difference near $180^o;$

 2) longitudinal component of magnetic field {\it in RW Aur A's wind}
becomes significant within a certain interval of the phase curve, i.e. at a
certain orientation of the system relative to the Earth, reaching the
maximum nearly 1/4 of the period after the passage of upper spot throug
the central meridian. It can be considered as an argument that
discovered magnetic field in the wind is related with a (quasi)stationary
outflow rather than sporadic gas ejection.

\begin{figure}[h!]
 \begin{center}
 \resizebox{10.0cm}{!}{\includegraphics{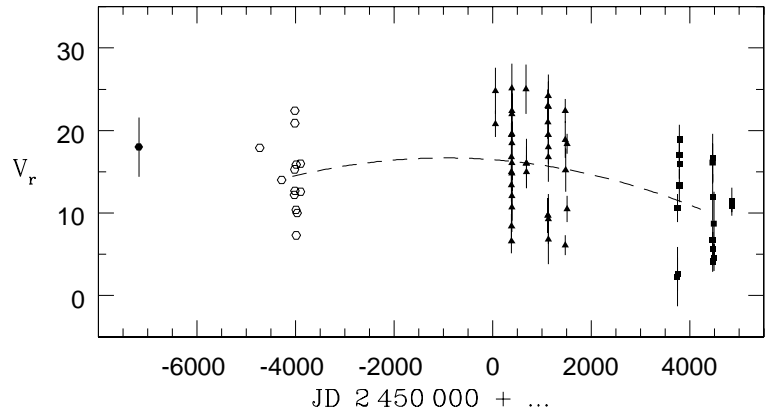} }
 \caption{Variations of radial velocities of RW Aur photospheric lines.
Single measurement of Herbig (1977) is depicted by filled circle and
measurements of Hartmann et al. (1986), Petrov et al. (2001) and our data
are shown with open circles, triangles and squares respectively.
The dashed line is drawn through average values of each subset.
}
\label{VrPhot}
  \end{center}
\end{figure}

  We found that average value of RW Aur A's photospheric lines radial
velocity $V_r^{ph}$ is smaller by 5 km\,s$^{-1}$ than value found by Petrov
et al. (2001). This discrepancy is seemed to be significant because the
amplitude of $V_r^{ph}$ variations due to stellar rotation is $\simeq 6$
km\,s$^{-1}.$ We have found two papers which contains results of more old
measurements of RW Aur's $V_r^{ph}$ measurements: the paper of Herbig (1977),
which contains only one measurement, and the paper of Hartmann et al. (1986),
which includes 13 measurements close in time. All available data are shown
on Fig.~\ref{VrPhot}. Parabola was drawn throught average values of
Hartmann et al., Petrov et.al and our data (dashed line).

			%%%%%%%%%%%%%%%%%%%%%
			%%%%%%%%%%%%%%%%%%%%%

  The measurements presented on the figure are insufficient for an
unambiguous conclusion about secular variations of RW Aur's radial velocity
and additional observations are necessary. However one can conclude
that variations of $V_r^{ph}$ by a few km\,s$^{-1}$ during $\sim 10$ years
cannot be explained by RW Aur B that located at a distance more than 100
a.u. from the primary. Thus we return again, but on different grounds than
Petrov et al. (2001), to a possibility that RW Aur is a triple system with
orbital period of hypothetic RW Aur C of several decades.

\bigskip

 We thank S. Alencar and P.\,P. Petrov for the spectra provided to
us as well as D.\,O. Kudryavtsev and D.\,A. Smirnov for the help with
observations.

			%%%%%%%%%%%%%%%%%%%%%%%%%%%%%%%%%%

\end{document}